%% file: ttk.tex
\documentclass[conference]{IEEEtran}
\ifCLASSINFOpdf
\else
\fi
\usepackage{url}
\usepackage{amsmath}
\usepackage{xcolor}
\usepackage{graphicx}
\usepackage{multirow}
\usepackage{amssymb}
\usepackage[caption=false,font=footnotesize]{subfig}
\usepackage[ruled,norelsize]{algorithm2e}
\SetKwComment{Comment}{$\triangleright$\ }{}

\makeatletter
\newcommand{\removelatexerror}{\let\@latex@error\@gobble}
\makeatother

\begin{document}
\title{Tracking Cyber Adversaries with Adaptive Indicators of Compromise}


 \author{

   \IEEEauthorblockN{Justin E. Doak (JD)\textsuperscript{+}, Joe B. Ingram, Sam A. Mulder, John H. Naegle, \\
     Jonathan A. Cox\textsuperscript{*}, James B. Aimone, Kevin R. Dixon, Conrad D. James}

   \IEEEauthorblockA{Sandia National Laboratories (SNL)\\
     Albuquerque, NM, USA \\
     Email: \{jedoak, jbingra, samulde, jhnaegl, jbaimon, krdixon, cdjame\}@sandia.gov \\
     \textsuperscript{+}Contact Author, \textsuperscript{*}Present Address: Netradyne, San Diego, CA, USA joncox@alum.mit.edu}

   \and
   
   \IEEEauthorblockN{David R. Follett}

   \IEEEauthorblockA{Lewis Rhodes Laboratories \\
     West Concord, MA, USA \\
     Email: drfollett@earthlink.net}

 }


%Telephone: (800) 555--1212, Fax: (888) 555--1212}


\maketitle

\begin{abstract}

A forensics investigation after a breach often uncovers network and
host indicators of compromise (IOCs) that can be deployed to sensors
to allow early detection of the adversary in the future.  Over time,
the adversary will change tactics, techniques, and procedures (TTPs),
which will also change the data generated.  If the IOCs are not kept
up-to-date with the adversary's new TTPs, the adversary will no longer
be detected once all of the IOCs become invalid.  Tracking the Known
(TTK) is the problem of keeping IOCs, in this case regular
expressions (regexes), up-to-date with a dynamic adversary.
%Two datasets, one containing data generated by the adversary and one
% containing data not generated by the adversary, are created with the
% intent to find the shortest regex that matches everything in the
% adversary dataset and nothing in the non-adversary dataset.  
Our framework solves the TTK problem in an automated, cyclic fashion
to bracket\footnote{``Bracket'' in this context means to contain a
  known adversary.  We contain him or her by keeping deployed IOCs
  up-to-date with his or her current TTPs.}  a previously discovered
adversary. This tracking is accomplished through a data-driven
approach of self-adapting a given model based on its own detection
capabilities.
  
In our initial experiments, we found that the true positive rate (TPR)
of the adaptive solution degrades much less significantly over time
than the na\"ive solution, suggesting that self-updating the model
allows the continued detection of positives (i.e., adversaries).  The
cost for this performance is in the false positive rate (FPR), which
increases over time for the adaptive solution, but remains constant
for the na\"ive solution. However, the difference in overall detection
performance, as measured by the area under the curve (AUC), between
the two methods is negligible.  This result suggests that
self-updating the model over time should be done in practice to
continue to detect known, evolving adversaries.


Full/Regular Research Papers, CSCI-ISCW

\end{abstract}



% \fi
%\IEEEpeerreviewmaketitle

\section{Introduction}
Once a corporate or government entity discovers that its network has
been compromised, a response team begins an investigation to respond
to the breach.  As part of the investigation, heavy use is made of
memory, disk, and network forensic capabilities.  The cyber security
analysts look for IOCs, such as IP addresses, domain names, file
hashes, and Windows Registry keys, that will allow them to
automatically detect the adversary in the future.  Additionally, the analyst
may create other more advanced signatures, such as regular expressions, to detect the
adversary. As adversaries
change their TTPs, some of the IOCs lose their effectiveness.  If all
of the IOCs become invalid, the adversary will be able to engage in
malicious activities without being detected.  It is likely that the
adversary will be rediscovered after a future breach and ensuing
incident response and then this process of extracting and deploying
IOCs to sensors is repeated.  A better approach to this problem would
be to automatically adapt the IOCs so that known adversaries can
continue to be tracked as they change their TTPs and the data they
generate correspondingly drifts over time.

The TTK problem is one of the fundamental challenges in cyber
security.  The complex adversary and defender dynamics that are
inherent in this problem can be modeled as a non-cooperative game in
game theory \cite{do2017game}.  Adversaries are aware that
defenders have deployed signatures to sensors to detect their activity, and
this leads them to constantly change TTPs such as command and
control (C\&C) messaging, the IPs and domains of servers used in C\&C,
and exploits directed at discovered vulnerabilities. In addition, many
of these indicators are extremely brittle, which makes it easy for the
adversary to intentionally manipulate his or her footprint to avoid
detection.  For example, if an IOC is a key in a Registry entry, the
adversary simply needs to write to a different location and then the
indicator will be invalidated. %no longer functions as an early warning system. 
If the indicator is the
IP address of a server used in C\&C communication, the adversary
simply needs to use another IP address and the IOC is no longer
valid. Similarly, the defender is also aware that the adversaries are
constantly changing their tactics to avoid detection. Manually updating
the IOCs to continue to bracket the adversary as behavior changes is
not possible given current resource allocations. %In this work, a
%methodology was developed to automatically keep IOCs current in order
%to bracket a continually evolving adversary. 

In this work, we describe a framework for solving the Tracking the Known problem. 
Starting with a base model of indicators, our framework automatically updates the
current model based on its predicted labels on the data stream. 
The intent is to self-adapt the model to concept drift in the data stream via a data-driven approach. In cyber defense,
analysts often rely on regular expressions as indicators to detect adversaries. A regular expression is a sequence of
characters that concisely represents a search pattern.  
Therefore, as an initial demonstration, we explore inducing regular expressions from
labeled data. Many such algorithms have been provided for learning
regular expressions due to the Regex Golf problem
\cite{1313:_Regex_Golf}: given two datasets of strings, create the
shortest regular expression that matches all the strings in one
set while not matching any of the strings in the other.

\section{Related Work}

To the best of our knowledge, there has been no prior work on the TTK
problem.  However, the notion of adapting or updating a model based on
its own outcomes is known as self-training in the semi-supervised
learning literature \cite{ChSc06}.

Regular expressions can be used to \emph{express} a regular language,
which is a formal language in theoretical computer science.  Finite
automata can be used to \emph{recognize} regular languages.  In fact,
regular expressions and finite automata are known to be equivalent by
Kleene's theorem \cite{kleene1951representation}.
%Thus, we include a brief discussion
%of some of the finite automata literature that may be relevant to our
%problem. 
Argyros et al. inferred web application firewall filters (i.e.,
regular expressions) by generating queries and observing the
responses.  They showed an improvement over existing automata learning
algorithms by reducing the number of required queries by $15\times$
through the use of symbolic representations \cite{argyros2016back}.
This has obvious overlap with Regex Golf since the queries that are
filtered can be considered part of the adversary dataset and those
that are passed can be considered part of the non-adversary dataset.
Prasse et al. attempted to infer a regular expression that matched a
given set of strings (i.e., email messages) and was as close as
possible to the regular expression that a human expert would have
created.  They applied their method to the problem of identifying spam
email messages.  Their technique frequently predicted the exact
regular expression a human expert would have created or the predicted
regular expression was accepted by the expert with little modification
\cite{prasse2015learning}.  Becchi et al.  showed how finite automata
can be extended to accommodate Perl-Compatible Regular Expressions (PCRE) \cite{becchi2008extending}.

Regular expressions have been used in many applications to solve
challenging cyber security problems.  Micron Technology, Inc. built a
massively parallel semiconductor architecture that directly
implemented regular expressions in hardware
\cite{dlugosch2014efficient}. The Machine Learning Lab at the
University of Trieste \cite{Machine_Learning_Lab} recently used
genetic programming to learn regular expressions
\cite{bartoli2014playing}.  The Norvig solution
\cite{Norvigs_Blog_Part_II} discussed in Section \ref{sec:regex} was
used as their baseline for comparison.
%The Norvig solution was used to generate regular expressions in the TTK pipeline.

\section{Tracking the Known}
The TTK problem is essentially a decision-theoretic problem on a data stream $D$. Given $x \thicksim D$, where $x$ represents some cyber datum/event (e.g., an HTTP session, Windows Registry key, IP address), the goal is to use a model to determine if $x$ was generated by an adversary or not, which is typically labeled by $y \in \{0, 1\}$, where $y = 1$ denotes the class of interest (i.e., the positive class). However, if the model is not updated against the stream, its performance will degrade over time as the data stream $D$ drifts.

Algorithm~\ref{alg:ttk} shows the process for updating the model in the TTK framework. For a given window $w$, the current model is used to label an event $x$ and then $x$ is added to the appropriate set ($P$ for a positive prediction and $N$ for a negative). At the end of the window, an algorithm $A$ is used to derive a new model $\hat{m}$, which is then concatenated with the current model in an ensemble-like fashion. (Some algorithms might update $m$ directly.) Then, the process is repeated on a new window. 

For the use case considered in this paper, the model is a list of
regular expressions that attempts to match on the positive class,
potentially indicating the presence of adversarial activity. However,
the framework is generic and, for instance, the model could be induced
using other machine learning algorithms. Basically, any algorithm that
can categorize $x$ as being generated by the adversary or not could be
utilized by this framework.

\begin{figure}[t]
  \centering
  \includegraphics[width=.9\linewidth]{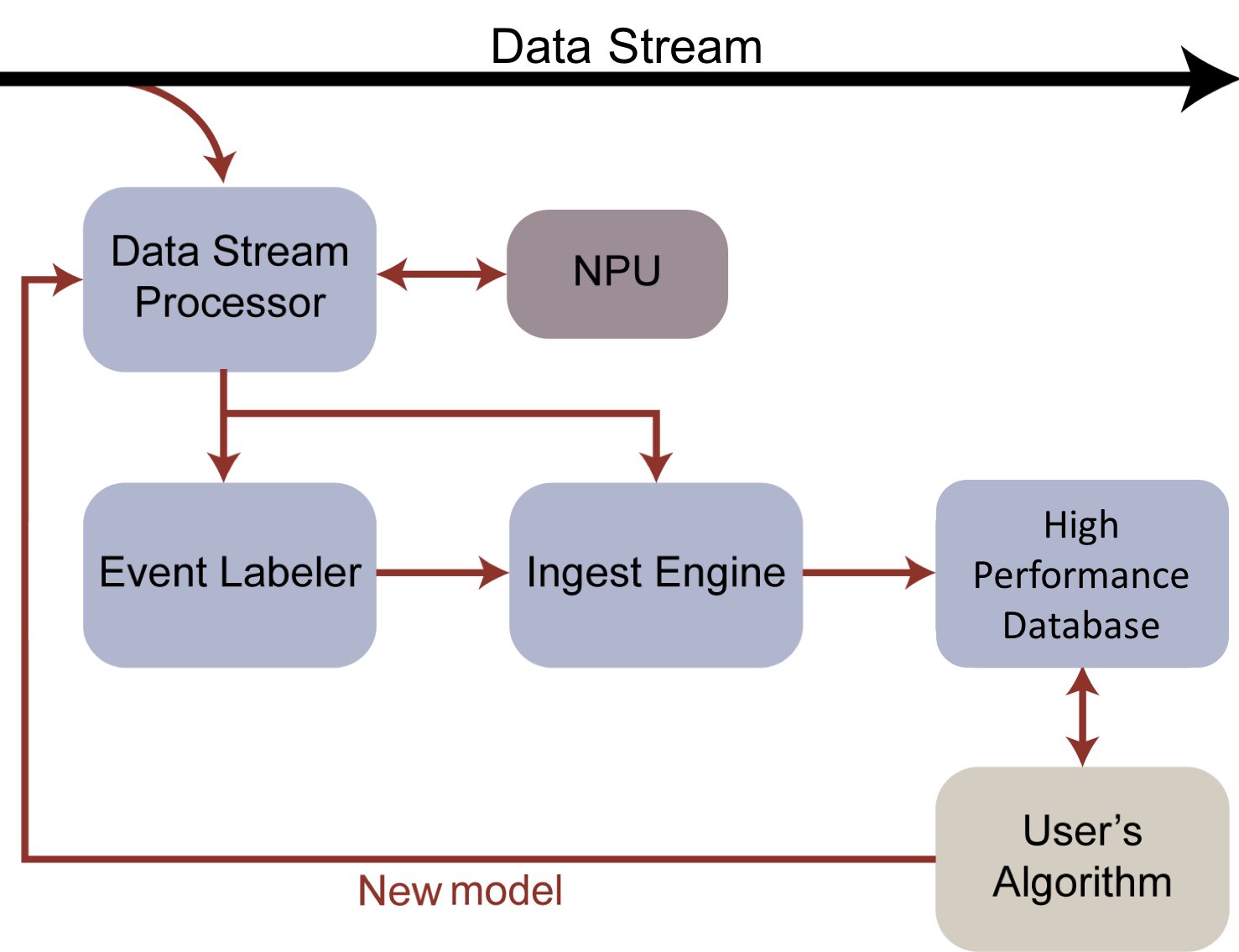}
  \caption{High-level View of the Implementation of the TTK Framework}
  \label{fig:ttk_testbed}
\end{figure}

\begin{figure}[!t]
 \removelatexerror
  \begin{algorithm}[H]
   \caption{Algorithm for Self-Updating Existing Model}
 \textbf{Input:} Current model, $m$; Window size, $w$;\\ \ \ \ \ \ \ \ \ \ Data stream, $D$; Algorithm, $A$ \\
 \textbf{Output:} Updated model: $\hat{m}$ \\
 $P = \{\}$ \\
 $N = \{\}$ \\
 \For{$i = 1$ to $w$} 
 {
  $x \thicksim D$ \Comment*[f]{Draw event from stream} \\
  $\hat{y} = m(x)$ \Comment*[f]{Get model's prediction} \\
  \eIf{$\hat{y} == 1$}{
     \Comment*[h]{Add event to positive set} 
    $P = P\ \cup\ \{x\}$
     }{ 
     \Comment*[h]{Add event to negative set} 
     $N = N\ \cup\ \{x\}$ 
  }
  }
  $\hat{m} = m\ \cup\ A(P, N)$  \Comment*[f]{Update model} \\
 \Return{$\hat{m}$}
  \label{alg:ttk}
  \end{algorithm}
\end{figure}

In order to solve the TTK problem, we created an automated, cyclic
pipeline to keep models of IOCs up-to-date with the data generated by
an adversary.  \figurename~\ref{fig:ttk_testbed} shows a high-level view
of the implementation of our system.  The cyclic nature of the pipeline can be described by
the following stages:

\subsubsection{Collect Data}

As cyber data is extremely voluminous, it must be captured, processed,
and stored in a near real-time, streaming manner. Therefore, a Data
Stream Processor must be efficient enough to process streaming cyber
data without losing events. Additionally, custom, high-performance,
neuromorphic hardware was utilized to further increase performance
where possible.

The Event Labeler is used to warm-start the detection system by
annotating events based on alternative mechanisms, such as a threat
feed or an analyst's detection rules. Additionally, a
human-in-the-loop might also be employed to relabel events and adapt
the system.

\subsubsection{Ingest Data}

A high-speed Ingest Engine was developed to ingest the processed cyber
data into a High-Performance Database. This database stores historical
cyber events that have been annotated. This historical data allows
models to be developed and validated before deployment.

\subsubsection{Partition Data}

Depending on the task of interest, a user can partition the data as
needed and derive a model based on that data.

\subsubsection{Generate Model}

The model can be induced from any algorithm that can discriminate
between two or more sets, such as supervised machine learning.

\subsubsection{Validate Model}

Before being deployed to the production system, the model is validated
against the historical, annotated data. Standard statistical machine
learning performance metrics are calculated (e.g., true positive rate,
false positive rate, etc.). If the model performs well historically,
it is deployed to the production system.

\subsubsection{Deploy Model}

During this stage, the new model is deployed and used by the Data
Stream Processor to process and annotate incoming cyber data.

\section{Learning Regular Expressions}
\label{sec:regex}

In this work, our approach to solving TTK requires an algorithm
for learning regular expressions: given two datasets of strings, create the
shortest regular expression that matches all the strings in one
dataset while not matching any of the strings in the other dataset.
As long as any of the IOCs remain valid, our solution continues to
track the adversary and hence identify the data that he or she
produces.  This data goes in one dataset while data not generated by
the adversary goes in another. The algorithm for deriving regular expressions is then repeated to
identify a more robust set of regular expressions for bracketing the
adversary and then those new, more current, rules are deployed.

\input{regex_golf_model}

\section{Case Study: Self-Adaptive Block List}

In order to adequately test the TTK framework, we selected a common
cyber security problem that could be solved by our framework, collected
and processed data for an extended period of time, and performed
experiments to validate our solution.  A common task in network
defense involves developing and deploying rules to detect and block
potentially malicious network traffic. Therefore, we selected this
problem as an initial case study.

\subsection{Collect Data}

The raw data are the ingress and egress packets collected at a
corporate network border.  The majority of this data are Hypertext
Transfer Protocol (HTTP) and HTTP over Transport Layer Security
(HTTPS) flows initiated by users on the inside of the network as they
visit various sites on the internet.  A streaming analysis tool was
used to decode the HTTP flows and extract the metadata/features.

Note that the Event Labeler attached the bootstrapping labels from a
blacklist to the HTTP flows.  After these labels were attached, the
initial adversary and non-adversary datasets were created from which
the initial regular expressions were derived.  The Event Labeler was
no longer needed at this point as future labels were provided by the
regular expressions themselves.

%For initial analysis, only the domain name was used.
%We add a feature, \emph{PCRE}, to contain the labels for
%the regular expressions that matched the data.

%For our experiments, the deployed regular expressions were applied to the
%domain name.  The regular expressions could be applied to other individual
%features or sets of features as well.  

%\subsubsection{Hardware Acceleration}
%\label{sec:dpu}

Two different methods were developed to process the regular expression
matching.  For ease of initial development of the overall system, the
WaterSlide open source software tool was used, which is a modular
metadata processing engine that is highly optimized for processing
streaming data \cite{WaterSlide}.  WaterSlide has a module that uses
the standard \texttt{re2} library, a highly optimized regular
expression engine, for processing regular expressions \cite{Re2}.
While the \texttt{re2} library is very efficient, processing complex
regular expressions is computationally intensive and represented a
significant performance bottleneck for the system.

\begin{figure}[t]
  \includegraphics[width=\linewidth]{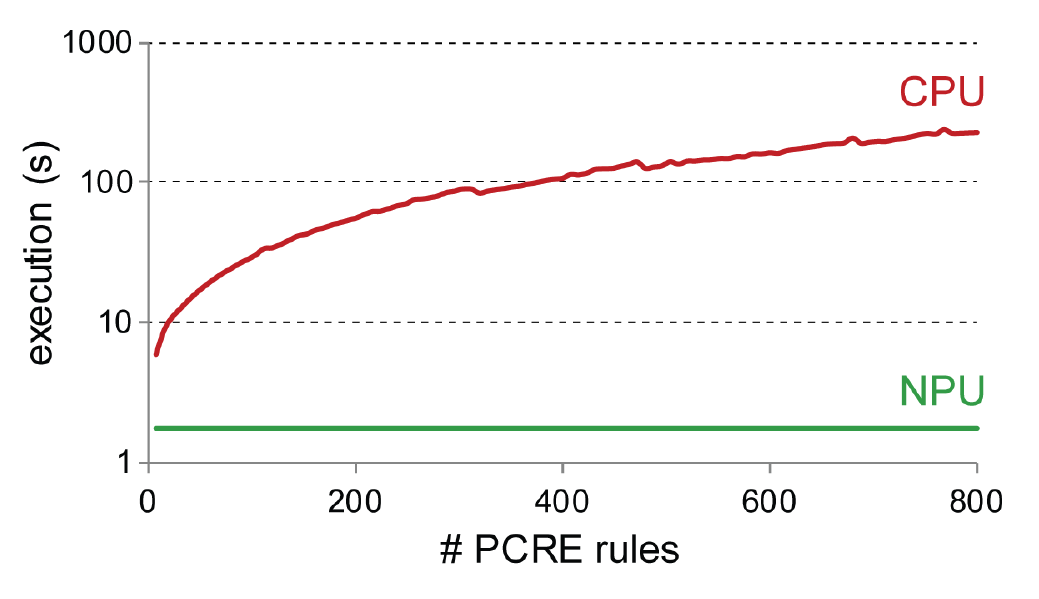}
  \caption{Execution Time as Regular Expressions are
    Added for CPU and NPU}
  \label{fig:DPU_plot}
\end{figure}

As part of the research to develop the TTK system, a module for
WaterSlide was developed that utilizes an FPGA-based regular
expression processing accelerator, i.e., the Neural Processing Unit
(NPU) \cite{follett2017neuromorphic}.  The architecture for the NPU
was motivated by the latest understanding of neuromorphic processing
models and greatly accelerated the processing of many regular
expressions as demonstrated in \figurename~\ref{fig:DPU_plot}.  The
implementation of the NPU was a PCI attached FPGA development board.
The NPU API was transparently implemented, completely hiding the
complexity of the NPU in a generally available library.

%A chip known as a Neural
%Processing Unit (NPU), which is designed
%to efficiently apply regular expressions to streaming data, was used.  The NPU
%provided a C API that we used to compile regular expressions and also
%to apply regular expressions to the streaming data.  Each regular expression was
%associated with a label that could represent, for example, the name of
%a specific adversary.  %All labels associated with regular expressions that matched
%the metadata in a flow are attached to that flow.

\subsection{Ingest Data}
 
The output of the data collection, the HTTP metadata and the labels of
any regular expressions that matched, was loaded into a database.  One
of the challenges was how to seed or bootstrap the system to create
the initial set of regular expressions.  Recall that our approach
relied on a dataset representing adversarial activity and a dataset
representing non-adversarial activity to generate the regular
expressions.  Initially, there were no deployed regular expressions,
hence they could not be used to split the incoming stream into the two
datasets.  To create the initial datasets, a blacklist
\cite{URLBlacklist.com} was used to classify the various domains into,
e.g., ads, news, e-commerce, banking, and dating after the HTTP data
had been loaded into the database. In an actual cyber security
context, these initial labels (or more generally IOCs) would have been
extracted by cyber security analysts in the context of a forensics
investigation into a breach.

The output of the Data Stream Processor is a set of tab-separated
values (TSV) files containing the HTTP metadata and the labels of any
regular expressions that matched.  We developed an ingest engine to
read the TSV files and then, using the Apache Phoenix
\cite{Apache_Phoenix} API, loaded those into a database, Apache HBase
\cite{Apache_HBase}.

\subsection{Partition Data}

Our solution required that the incoming HTTP flows be split into the
dataset of interest and the dataset not of interest, based on the
specific task that an analyst might want to automate.
%For our initial experiments, all data from domains whose primary category was ``ads''
%according to the blacklist were placed into the dataset of interest.
%All data from domains whose primary category was something other than
%ads were placed in the dataset not of interest.  
Then, a set of regular expressions to detect the dataset of interest
was derived and
%on these two datasets to create the initial regular expressions 
deployed to the sensor.  After deployment, it was the regular
expressions themselves, not the bootstrapping labels from the
blacklist, that bifurcated the incoming flows into the appropriate
datasets.

\subsection{Generate Model}

Once the data has been split into the two datasets, we play Regex Golf
to generate the more up-to-date regular expressions to bracket the
threat.  Our implementation is based on a solution by Peter Norvig
\cite{Norvigs_Blog_Part_I, Norvigs_Blog_Part_II}.  Norvig's solution
requires that the datasets be disjoint, i.e., no string can appear in
both the adversary and the non-adversary datasets.  This assumption
may be problematic in the cyber security context.  For example, a
network sensor may detect a downloaded file that could appear both in
conjunction with a malware toolset but also occur normally as part of
a legitimate toolset.  A good example of this is the \texttt{netcat} program,
which is often an indicator of malware, but is also a legitimate
sysadmin tool.  In fact, many antivirus products will flag a zip file
if it contains \texttt{netcat} as potential malware.  Norvig's algorithm will
also always find a perfect solution, i.e., it will find a set of
regular expressions that match everything in the adversary dataset and
nothing in the non-adversary dataset.  Relaxing these two constraints
is an area of future work that might lead to an algorithm that is
better-suited for deployment.  Section~\ref{sec:regex} provides more
detail on this algorithm.

\subsection{Validate Model}

A method to validate regular expressions after they were generated by
the Regex Golf model was developed.  First, a vector of tuples for
each dataset showing the true class of each string (e.g., adversary)
and the class predicted by the algorithm was created.  These vectors
were passed to the \texttt{scikit-learn} library \cite{scikit-learn}
to calculate standard metrics such as precision, recall, accuracy,
etc.  Given the limitations of the current algorithm, these metrics
are all currently perfect.  However, if we relax the perfect
classification constraint, these metrics could become meaningful.

\subsection{Deploy Model}

The regular expressions learned by the Regex Golf model were written
to a file, along with appropriate labels, after the generation and
validation processes.  The streaming analysis engine recognized that
the file had been modified, which triggered a reload of that file and
its associated regular expressions and labels.  These regular
expressions were compiled by the NPU and then the NPU was used to
accelerate application of the regular expressions to the network data.
The system has now returned to the beginning of the pipeline and data
was collected using these newly-deployed regular expressions.  This
created new adversary and non-adversary datasets and the automated,
cyclic process continued.

\begin{figure*}[t!]
  \centering
  \includegraphics[width=.75\linewidth]{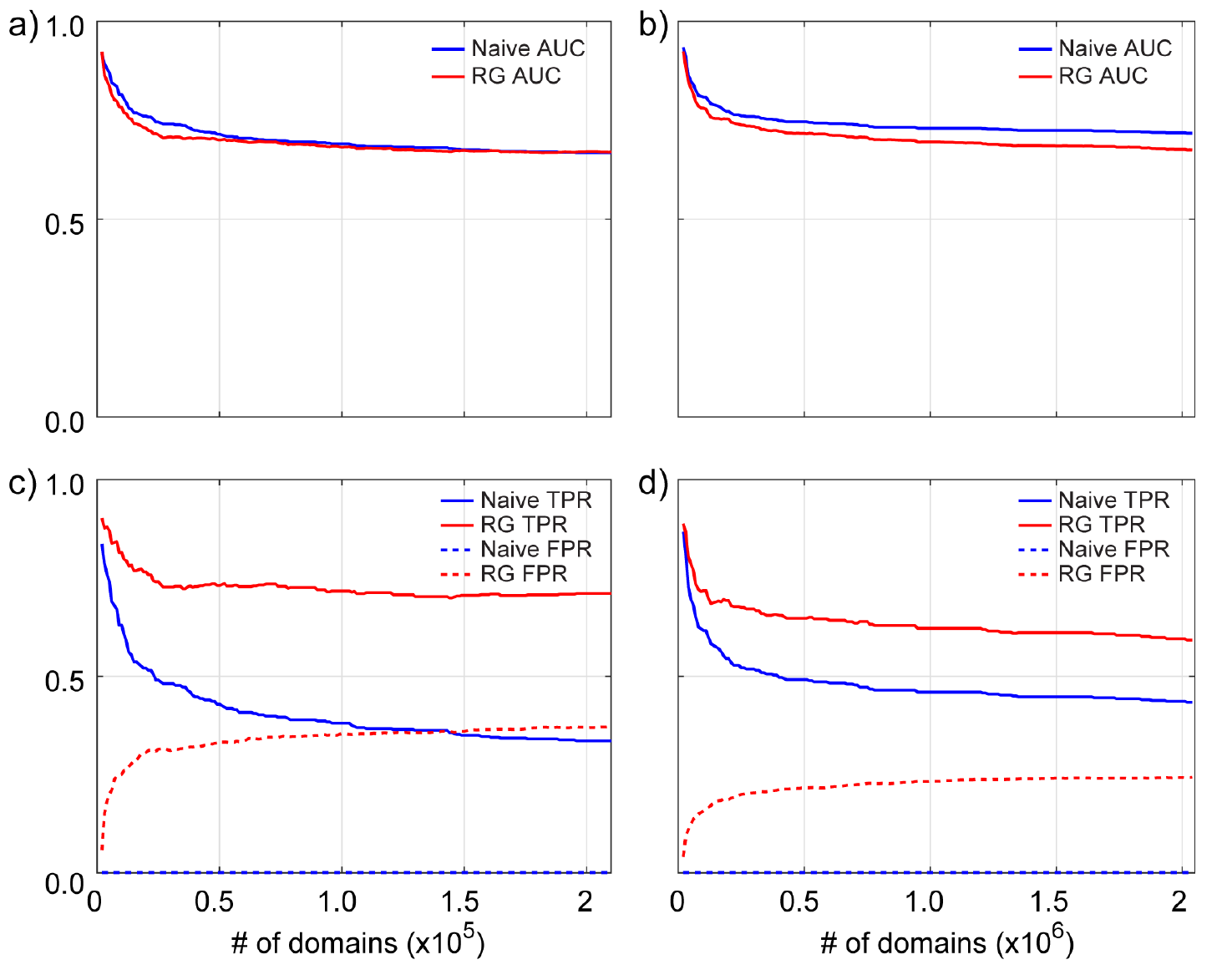}
  \caption{AUC for a) window size 1,000 and b) window size
    10,000. TPR/FPR for c) window size 1,000 and d) window size
    10,000.}
  \label{fig:perf}
\end{figure*}

\section{Experimental Results}

As an initial experiment, we investigated the problem of tracking
advertising domains. Data was collected at a corporate network border
for two weeks, resulting in approximately 23 million unique HTTP
flows. Of those flows, roughly 34\% were identified as advertising
domains by the list of domain categories, which was used as ground
truth.

The TTK problem is analogous to finding positive examples in a binary
classification task.  Therefore, some common metrics for estimating
classification performance were used.  The true positive rate (TPR) is
the fraction of positive examples that were correctly identified by
the model and is represented as $\text{TPR} = \text{TP} / \text{P}$
where TP is the number of positives correctly classified and P is the
total number of positive samples.  The false positive rate (FPR) is
the fraction of negative examples (non-advertising domains) that were
identified as advertising domains.  It is defined as $\text{FPR} =
\text{FP} / \text{N}$, where FP is the number of negatives incorrectly
classified and N is the total number of negative samples.  A common
metric for summarizing the detection performance is the Receiver
Operating Characteristic (ROC) curve, which plots the FPR (on the
$x$-axis) against the TPR (on the $y$-axis). Computing the area under
the ROC curve (AUC) provides a single metric for summarizing detection
performance. An AUC value of 1.0 represents perfect detection
performance, while an AUC score of 0.5 means that the model does no
better than random.

\figurename~\ref{fig:perf} shows the cumulative performance for window sizes
of 1,000 and 10,000. The window size indicates how many domains are used
to seed the models, as well as how often metrics are collected.
%(Rolling in this context means that the
%algorithms are tested on the domains in the current window and all
%previous windows except the first.)  
The na\"ive solution is simply
the list of advertising domains seen in the first window and is
analogous to a standard blocklist that only utilizes domain names.
For the na\"ive solution, the FPR did not change from 0.0 because our
ground truth doesn't change.  In other words, the same domains that
were marked ads in the beginning are still labeled ads as we continue
the experiment.  We also see that for the na\"ive solution the TPR
went down because new domains were seen that were in the advertising
category, but weren't in the original list of advertising domains used
for detection.

For the Regex Golf solution, the FPR went up because the learned
regular expression marked domain names as advertisements that were
not. This increase is an unfortunate side effect of the self-training
paradigm in that errors tend to propagate.  On the other hand, the TPR
remained steady or decreased only slightly because the learned regular
expression had some generalizability and could correctly recognize
some previously unseen domains as advertisements.
  
For the smaller window size, the na\"ive solution shows a 60\%
decrease in ability to detect positives, while the Regex Golf solution
only shows a 22\% decrease over time with a negligible difference in
overall detection performance.  For the larger window size, the
na\"ive solution shows a 50\% decrease in ability to detect positives,
while the Regex Golf solution only shows a 33\% decrease over time.
The Regex Golf solution has a 6\% lower overall detection performance
than the na\"ive solution.  In summary, it is apparent that the Regex
Golf solution is finding more positive instances with only a slight
degradation in performance, as indicated by the slightly lower AUC
scores.

%Note the increased variance present in the windowed results.  Each
%point in the windowed results corresponds to calculating the metric on
%the next 1,000 domains.  Hence, you are likely to get significant
%fluctuation between each new 1,000 domains that the model tries to
%classify.  On the other hand, the rolling approach is cumulative and
%classifies not only the next 1,000 domains, but also all previous
%domains after the first 1,000 used for training.

\section{Conclusions and Future Work}

In this work, we demonstrated the ability to keep IOCs (i.e., regular
expressions) up-to-date with an evolving adversary.  This will allow
network sensors to continue to bracket these existing threats as they
change their TTPs.  Not losing the ability to detect these adversaries
will save countless hours in incident response because they will be
identified before they have breached our networks or at least before
they have spread laterally.  It will also prevent these adversaries
from accomplishing their objectives, e.g., exfiltrating desired
information.

The current solution for generating regular expressions requires that
the two datasets are disjoint. This constraint is not realistic to
impose in a production cyber security context.  In addition, the
current model will always find a perfect solution; as the datasets
grow larger, this may prevent the model from ever converging.
Relaxing this constraint could lead to much faster build times, again
advantageous in a production cyber security context.

Currently, our system is only utilizing a single feature: the domain
name. By utilizing other features of HTTP traffic, we anticipate even
better results.  Given the performance increase provided by the NPU,
deploying a larger set of regular expressions that match on different
features is feasible, perhaps by using a custom set multicover
optimization.  Additionally, our framework allows for more complex
models to be employed. Any algorithm that can partition data or solve
a learning task can be supported by the framework. Therefore, trying
more advanced methods, such as supervised learning or deep learning,
is also an area of future work. Deep learning can also be used in the
Data Stream Processor to transform features into more appropriate
representations for learning.

Finally, we would like to deploy this in cyber security operations
using the regular expressions discovered in a forensics investigation
as the first set of deployed regular expressions.  We will then be
able to quantify how well the system brackets an adversary and their
changing TTPs over time.
\section*{Acknowledgment}

%% The authors received support from SNL's LDRD Program, and specifically
%% the Hardware Acceleration of Adaptive Neural Algorithms Grand
%% Challenge Project. SNL is a multimission laboratory managed and
%% operated by National Technology and Engineering Solutions of Sandia,
%% LLC., a wholly owned subsidiary of Honeywell International, Inc., for
%% the U.S. DOE's NNSA under contract DE-NA0003525.

The authors acknowledge financial support from Sandia National
Laboratories' Laboratory Directed Research and Development Program,
and specifically the Hardware Acceleration of Adaptive Neural
Algorithms (HAANA) Grand Challenge Project. Sandia National
Laboratories is a multimission laboratory managed and operated by
National Technology and Engineering Solutions of Sandia, LLC., a
wholly owned subsidiary of Honeywell International, Inc., for the
U.S. Department of Energy's National Nuclear Security Administration
under contract DE-NA0003525.

% \IEEEtriggeratref{9}


%\bibliographystyle{IEEEtran}
% \end{thebibliography}

\input{bibliography}

\end{document}

%% file: regex_golf_model.tex
If it is assumed that it is possible to create two distinct sets as
required by Norvig's algorithm, there is a straightforward solution:
combine the full contents of the adversary dataset using logical
disjunction (i.e., the $OR$ operator) to create a very long regular
expression that provides a perfect solution.  Of course, this solution
is unwieldy, so we want to optimize the size of the regular expression
required to separate the two sets.  A regular expression to do this
will consist of some set of matching conditions $OR$-ed together.
This problem can be generalized as the set cover problem, usually
stated as:

Given a set of elements $U = {1, 2, ..., n}$ and a collection $S$ of
$m$ sets whose union equals $U$, identify the smallest sub-collection
of $S$ whose union equals $U$.

As an example, consider $U$ as all integers from $1$ $to$ $10$,\\ and
$S = \{ \{1, 2\}, \{2, 3, 4, 5\}, \{2, 4, 6\}, \{4, 6, 8\}, \{1, 3,
5\},\\ \{7, 9\}, \{1, 10\} \}$.  The union of all sets in $S$ is $U$,
meaning $S$ covers $U$, but we can find a smaller sub-collection that
still has this property.  In this case, $\{ \{1, 10\}, \{2, 3, 4,
5\}, \{4, 6, 8\}, \{7, 9\} \}$ also covers $U$.

Optimizing to find the smallest collection that covers the original
set is an NP-hard problem, with the decision version\footnote{The
optimization problem involves finding a solution, while the decision
version involves determining if a solution exists.} of the problem
proved NP-complete in \cite{karp1972reducibility}.  Given this, the
best way to approach the problem is through approximation.

Norvig's algorithm begins by generating a set of regex components as
follows.  Each dataset is first broken into $n$-grams with $n$ ranging
in size from 1 to the length of the longest string in the dataset and
then a set of all possible subsequences of these sizes is created.
For this set, a '.' is then added in every possible iteration of these
subsequences replacing some number of characters.  Next, regular expression
components are created by inserting special repetition characters, such as '+',
'*', or '?', after each character that is not '.' in every possible
combination.  From this set, any component that matches anything in
the non-adversary dataset is filtered out.  This provides a set of
components to choose from that match at least one string in the
adversary set and no strings in the non-adversary set.

These components are ranked based on how many strings they match and
the best is added to a solution set.  The strings already covered are
removed from consideration, and the remaining components are again
ranked by how many of the remaining strings they match.  This process
is repeated until all strings are covered.  This is a purely greedy
algorithm.

%% file: bibliography.tex
%\IEEEtriggeratref{5}

% \bibliographystyle{./bib/IEEEtranS}
% \bibliography{./bib/IEEEabrv,./bib/ttk}
\bibliographystyle{./IEEEtranS}
\bibliography{./IEEEabrv,./ttk}